\documentstyle[12pt,epsf]{article}

\begin{document}
\begin{center}
ASYMPTOTIC BEHAVIOR OF NUCLEON ELECTROMAGNETIC FORM FACTORS IN SPACE- 
AND TIME-LIKE REGIONS
\vskip 5mm

Egle Tomasi-Gustafsson$^{(1)}$ and Michail P. Rekalo$^{(2)}$

\vskip 5mm

(1) {\it DAPNIA/SPhN, CEA/Saclay, 91191 Gif-sur-Yvette Cedex, 
France}

\vskip 3mm

(2) {\it Middle East Technical University, 
Physics Department, Ankara 06531, Turkey

and

National Science Center KFTI, 310108 Kharkov, Ukraine}

\vskip 5mm


\centerline{\bf Abstract}

\vskip 3mm

We compare the existing data about electromagnetic form factors of the proton 
in 
the time-like region of momentum transfer (up to $q^2\simeq$ 14 (GeV/c)$^2$),  
with the corresponding data in the space-like region. From 
the 
constrains given by the Phragm\`en-Lindel\"of theorem, it turns out that  the 
asymptotic regime can not be reached simultaneously for both form 
factors (electric and magnetic) in the considered region of momentum transfer.  
The 
angular dependence for  the annihilation 
processes, such as $\overline{p}+p\leftrightarrow e^++e^-$, is sensitive to the 
asymptotic properties of form factors and to the two-photon physics in the 
time-like 
region.
\end{center}
\vskip 10mm
Recent experimental data on nucleon electromagnetic form factors (FF) in 
time-like 
(TL) \cite{Am99} and space-like (SL) \cite{Jo00} regions of momentum transfer 
square $q^2$, and new theoretical 
developments 
\cite{Ra00} show the necessity of a global description of form 
factors in the full region of $q^2$. The question 
of "where the asymptotic behavior for form factors is reached" is 
often discussed in literature, however different expectations and predictions 
are given by different models (for a recent review, see \cite{Ra00}). Our aim is 
to discuss the experimental data 
about proton electromagnetic structure in SL and TL regions, 
to estimate (independently from current models)  where the asymptotic region is, 
and, finally, to suggest some observables to be measured in order to test these  
predictions.

For the description of the nucleon  electromagnetic structure in the full region 
of momentum transfer square (space-like and time-like), two form factors are 
defined, electric, $G_E$, and magnetic, $G_M$,
which are related to the Dirac ($F_1$) and Pauli ($F_2$) FFs  by: 
$$G_E=F+\tau F_2,~~G_M=F_1+F_2, ~~\tau=\displaystyle\frac{s}{4m^2},$$
where $m$ is the nucleon mass and  $s$ is the square of the total 
energy in $e^++e^-$ $(p+\overline{p})$ collisions, i.e. the time-like equivalent 
of the 
four-momentum transfer square $q^2$ for $e+p-$scattering.

In the QCD-approach, two aspects are related to the asymptotic region:
1) the  $q^2$-dependence of the electromagnetic hadronic form factors, in 
accordance with the quark counting rule \cite{Br73,Ma73} and 2) the 
conservation of hadron 
helicity \cite{Le80}. The existing experimental data about electromagnetic form 
factors of 
pion, nucleon and deuteron confirm the quark counting behavior: 
$F_A(q^2)\simeq 1/(q^2)^{n_A-1}$ where $n_A$  is the number 
of elementary 
constituents in the hadron $A$. On the other hand the hypothesis of 
helicity conservation, which constrains in particular polarization 
observables, does not work successfully, for the deuteron form 
factors \cite{Ab00} and especially for the electromagnetic form factors of 
$N\to\Delta$ 
transitions \cite{Fr99}, where the electric quadrupole (transversal and 
longitudinal) form 
factors, measured up to $q^2=$ 4 (GeV/c)$^2$, are very small, in evident 
contradiction with helicity conservation. This situation is 
confirmed also by the study of different electromagnetic processes: pion 
photoproduction, $\gamma+N\to N+\pi$ \cite{An76}, nucleon Compton scattering, 
$\gamma+p\to \gamma+p$ \cite{Sh79}, deuteron photodisintegration, 
$\gamma+d\to n+p$ 
\cite{Bo98} and coherent 
$\pi^0$-photoproduction, $\gamma+d\to d+\pi^0$ \cite{Me99}.
Note that the scaling behavior of the differential cross section is especially 
evident in elastic $pp-$scattering \cite{La73}, where the data are consistent 
with the scaling law predictions over a wide range of  angles and energies.

The comparison of the nucleon electromagnetic FF in SL and TL regions opens
the way to a general and model independent  discussion of asymptotic 
properties. Form factors are analytical functions of $q^2$, 
being real functions in the SL  region (due to the hermiticity of the 
electromagnetic Hamiltonian) and complex functions in the 
TL region. The Phr\`agmen-Lindel\"of 
theorem \cite{Ti39} gives a rigorous prescription for the asymptotic behavior of 
analytical functions: 
$\lim_{q^2\to -\infty} F^{(SL)}(q^2) =\lim_{q^2\to \infty} 
F^{(TL)}(q^2)$.
This means that, asymptotically, the FFs, have the following constrains: 
1) the time-like phase vanishes  and 2) the real part of the FFs, 
${\cal R}e  F^{(TL)}(s)$, coincides with the 
corresponding value $F^{(SL)}(q^2)$.

The existing experimental data about the electromagnetic FFs of charged pion or 
proton in the time-like region do not allow a similar complete test of   the 
Phr\`agmen-Lindel\"of theorem, especially concerning the vanishing phase. Even 
in the simplest case  of the  pion form 
factor, 
$F_{\pi}(s)$, its phase (which is the physical quantity fixed by the unitarity 
condition) can not be  derived from  the cross section of the 
$e^+ + e^-\to\pi^+ +\pi^-$ process, as it depends on $|F_{\pi}(s)|^2$, only.
This is also true for the proton FFs. However $T-odd$ polarization phenomena in 
$e^++e^-\leftrightarrow  \overline{p}+p$ are sensitive to the relative phase  
$\delta$ of the electric and 
magnetic FFs. The simplest polarization observables, which are characterized by 
the product ${\cal I}mG_E G_M^*\simeq \sin\delta$ are the  asymmetries in the 
reactions
$\vec {e^+ }+e^-\to \overline{p}+p$ and $\overline{p}+\vec p\to e^++e^-$.
However no polarization observables in these reactions allow to extract  the 
absolute value of the phase. Only the study of more complicated reactions such 
as $\pi^-+p\to n+e^++e^-$ \cite{Re66} or $\overline{p}+p\to \pi^0+e^++e^-$ 
\cite{Du96} allows, in 
principle: 

\noindent - to determine the nucleon FFs in the unphysical region of TL momentum 
transfer, 
for 4$m_e^2 \le s\le$ 4$m^2$, where $m_e$ is the leptonic mass;

\noindent - to determine the relative phase of pion and nucleon form factors.

In this work we focus on $e^-+p$-elastic scattering and on the processes 
$e^++e^-\leftrightarrow  \overline{p}+p$. We suggest here a new 
analysis of FFs in TL region and  discuss 
experimental methods to check the different assumptions.

The recent measurement, in TL region, of the experimental cross section  in the 
annihilation 
process $\overline{p}+p\to e^++e^-$ extends up to $s\simeq$ 14 (GeV/c)$^2$. 
The cross section can be expressed as a function of 
FFs according to the following formula \cite{Zi62}:
\begin{equation}
\displaystyle\frac{d\sigma}{d(cos\theta)}=
\displaystyle\frac{\pi\alpha^2}{8m^2\tau\sqrt{\tau(\tau-1)}}
\left [ \tau |G_M|^2(1+\cos^2\theta)+|G_E|^2\sin^2\theta
\right ],
\end{equation}
where $\theta$ is the angle between the electron and the antiproton
in the center of 
mass frame. 
The angular dependence of the differential cross section can be written as:
\begin{equation}
\displaystyle\frac{d\sigma}{d(\cos\theta)}=
\sigma_0\left [ 1+{\cal A} \cos^2\theta \right ],
\end{equation}
where $\sigma_0$ is the value of the differential cross section at 
$\theta=\pi/2$ and the  asymmetry ${\cal A}$,  can be expressed as 
a function of the FFs:
\begin{equation}
{\cal A}=\displaystyle\frac{\tau|G_M|^2-|G_E|^2}{\tau|G_M|^2+|G_E|^2}
\end{equation}

The Rosenbluth 
separation of $|G_E|^2$ and $|G_M|^2$ in TL region, has not been realized yet.

In order to extract the form factors,  due to the poor statistics, it 
is necessary to integrate the differential cross section over a wide angular 
range. One assumes that the $G_E$-contribution plays a minor role in the cross 
section at large $s$ and the 
experimental results are usually given 
in terms of $|G_M|$, under the hypothesis that $G_E=0$ (case 1) or $|G_E|=|G_M|$ 
(case 2). 
The second hypothesis is strictly valid at threshold only, but 
there is no 
theoretical argument which justifies its validity at any other momentum 
transfer, where $s\neq 4m^2$. The first hypothesis is arbitrary.

The $|G_M|^2$ values depend, in principle, on the kinematics where the 
meaurement was performed and the angular 
range of integration, however it turns out that these two assumptions for $G_E$ 
lead to comparable values for $|G_M|$.

In the SL region the situation is different. The cross section for the elastic 
scattering 
of electron on protons is sufficiently large to allow the measurements of 
angular 
distribution and/or of polarization observables. The existing data on 
$G_M$ show a dipole behavior  up to the highest 
measured value, $q^2\simeq$ 31 (GeV/c)$^2$ according to
\begin{equation}
G_M(q^2)/\mu_p=G_d,~{\mbox with}~
G_d=\displaystyle\frac{1}
{\left [1+\displaystyle\frac{q^2}{ m_d^2 }\right ]^2},~m_d^2=0.71~(GeV/c)^2.
\end{equation}

It should be noticed that the independent determination of both $G_M$ and $G_E$ 
FFs, from the unpolarized $e^-+p$-cross section, has been done up to $q^2=$ 8.7 
(GeV/c)$^2$ \cite{And94}, and the further extraction of $G_M$ 
\cite{Ar86} assumes $G_E=G_M/\mu_p$.
The behavior of $G_E$, deduced from polarization experiment $p(\vec e,e'\vec p)$ 
differs from $G_M/\mu_p$,
with a deviation from $G_d$ up to 50\% at $q^2$=3.5 (GeV/c)$^2$ \cite{Jo00}. 
This is the maximum momentum at 
which the new, precise data are available, which  corresponds to 
values of $s$ just under threshold of the 
reaction $\overline{p}+p\to e^+ +e^-$, when translated to TL region. The new 
data can still be fitted by a 
dipole function, also, but with a smaller value of $m_d^2=0.61$ (GeV/c)$^2$.

The experimental situation is summarized in Fig.1, where the data are 
normalized to the function $G_d$. 
The values of $G_M$ in the TL region, 
obtained under the assumption  that $|G_E|=|G_M|$, are larger than the 
corresponding SL values. This has been considered 
as a proof of the non applicability of the Phr\`agmen-Lindel\"of theorem, 
or as an evidence that  the asymptotic regime is not reached \cite{Bi93}.

The magnetic form factor of the proton in the TL region (which is deduced 
from the hypothesis $G_E=0$ or $G_E=G_M$), can be parametrized as: 
$G_{M}^{(TL)}=G_d\displaystyle\frac{a}{\left 
(1+\displaystyle\frac{s}{m_{nd}^2}\right)}$, where $a$ is a normalization 
parameter and $m_{nd}^2=3.6\pm 0.9$ (GeV/c)$^2$ characterizes the deviation 
from the usual dipole $s$-dependence. The extrapolation to high $q^2$ based on 
this formula (Fig. 1, full line), indicates that the 
Phr\`agmen-Lindel\"of theorem 
will be satisfied by this FF, only for $s(q^2)\ge 20$ (GeV/c)$^2$.

The value of the mass parameter $m_{nd}^2$ is comparable for other 
electromagnetic form factors: the electric FF of the proton ($m^2\simeq$ 5.3 
(GeV/c)$^2$) 
(Fig. 1, dashed line) and the magnetic FF of the $N\to \Delta$ transition 
($m^2\simeq$ 6.1 (GeV/c)$^2$). This might be an indication 
that this parameter 
is related to the internal hadronic electromagnetic structure.

Let us consider now another procedure for the extraction of FF in the TL 
region assuming that at least one of the 
two proton electromagnetic FFs  has reached the asymptotic regime. 
This looks as  a reasonable hypothesis for $G_M$, 
which shows an early scaling behavior, in accordance with quark counting rules.

In this case the Phragm\`en-Lindel\"of theorem constrains definitely $|G_M|$ in 
TL region to have the same value as in SL, and therefore from Eq. (1) we can 
deduce $|G_E|$, using  the existing 
experimental data about $\overline{p}+p\leftrightarrow e^+ +e^-$ (case 3). A 
fourth possibility is taking for 
$G_E$ in the TL region the values suggested from the SL region (i.e. assume that 
$G_E$ is asymptotical in the considered region), and calculate 
$|G_M|$ (case 4).

We report, in  Fig. 2, some of the recent  data in TL region, reanalized 
following the  possibilities suggested above. Fig. 2a shows the values 
of the form factors. For case 3, where $G_M$ is taken according to Eq. 4 ($G_M$ 
is asymptotical), the values of 
$|G_E|$ are plotted and they are larger than in cases 1 and 2. This seems 
to suggest that asymptotics are not reached for $G_E$, as the values in 
SL and TL gets more apart. 

On the other hand, taking for 
$|G_E|$ the SL values (case 4), affects very little the values of $G_M$, due to 
the kinematical factor 
$\tau$, which weights the magnetic contribution to the differential cross 
section.

Therefore, the existing data about $\overline{p}+p\to e^+ +e^-$ do not 
contradict the hypothesis that one  form factor (electric or magnetic) could be 
asymptotic at relatively large momentum transfer ($s\simeq 6\div 14$ 
(GeV/c)$^2$), but in this case the other form factor is far from the asymptotic 
regime. Although affected by large statistical errors, the existing angular 
distributions in TL region \cite{Ba94,Bi83} do not contradict the possibility 
that the electric and magnetic FF might differ substantially. Note also that the 
new data about $G_{Ep}$ in the SL region do 
not change essentially $|G_M|$ in the TL region, in comparison with the 
standard analysis (cases 1 and 2).

Fig. 2b shows the asymmetry for cases 3 and 4. Case 1 and case 2  give, 
respectively,  ${\cal A}=1$ and ${\cal  A}=(\tau-1)/(\tau+1)$.

The predicted asymmetry is very sensitive to the different underlying 
assumptions. 

The measurement of the differential 
cross section for the process $\overline{p}+p\to e^+ +e^-$ at a fixed value of 
$s$ 
and for two different angles $\theta$,  allowing  the separation of the two FFs, 
$|G_M|^2$ and $|G_E|^2$, is equivalent to the well known Rosenbluth separation 
for 
the elastic $ep$-scattering. However in TL, this procedure is simpler, as it 
requires to change only one kinematical variable, $\cos\theta$, whereas, in SL 
it is 
necessary to change simultaneously two kinematical variables: the energy of the 
initial electron and the electron scattering angle, fixing the momentum transfer 
square, $q^2$. 

The angular dependence of the cross section, Eq. (2), results 
directly from the assumption of one-photon exchange, where the spin of the 
photon 
is equal 1 and the electromagnetic hadron interaction satisfies the 
$C-$invariance. 
Therefore the measurement of the differential 
cross section at three angles would allow to test the presence of $2\gamma$ 
exchange also. The interference of C-odd amplitude of the one-photon exchange 
with 
the 
C-even amplitude of the two-photon exchange, will give rise to odd 
$\cos\theta-$terms in the cross section:
\begin{equation}
\displaystyle\frac{d\sigma}{d(\cos\theta)}(\overline{p}p\to e^+e^-)=
\sigma_0
\left [ 1+a_1\cos\theta+ a_2 \cos^2\theta + a_3 \cos^3\theta + ...\right ],~~
\end{equation}
where $a_i,~i=1,2,3 ..$, are $s-$dependent real coefficients.

In order to cancel the possible effects which can be induced by the two-photon 
contribution, one can suggest the following procedure:
\begin{itemize}
\item Consider the sum of the differential cross section at two angles, $\theta$ 
and $\pi-\theta$, i.e.:
\begin{equation}
\displaystyle\frac{d\sigma}{d\cos\theta}(\theta)+ 
\displaystyle\frac{d\sigma}{d\cos\theta}(\pi-\theta)=2
\sigma_0 \left [1+a_2 cos^2\theta...\right ]
\end{equation}
\item Solve the following integral in which the $\cos\theta$-odd terms 
disappear:
\begin{equation}
\int_{-\cos\theta_{max}}^{\cos\theta_{max}}
\displaystyle\frac{d\sigma}{d\cos\theta}(\theta)d\cos\theta
=2\sigma_0\cos\theta_{max}\left [1+3a_2\cos^2\theta_{max}+..\right ]
\end{equation}
\end{itemize}
On the other hand the following difference is very sensitive to the 
$1\gamma\times 
2\gamma$-interference contribution:
\begin{equation}
\displaystyle\frac{d\sigma}{\cos\theta}(\theta)- 
\displaystyle\frac{d\sigma}{d\cos\theta}(\pi-\theta)=
2\sigma_0 \left [ a_1 cos\theta +a_3 cos^3\theta...\right ]
\end{equation}
The integration gives:
\begin{equation}
\int_{0}^{\cos\theta_{max}}
\displaystyle\frac{d\sigma}{d\cos\theta}(\theta)d\cos\theta-
\int^{0} _{-\cos\theta_{max}}
\displaystyle\frac{d\sigma}{d\cos\theta}(\theta)d\cos\theta-
=2\sigma_0 \cos\theta_{max}\left [
a_1  +\displaystyle\frac{1}{2}a_3 cos^2\theta_{max}+..\right ]
\end{equation}

A similar procedure can be suggested in the SL region, through the comparison of 
electron and positron scattering in the same kinematical conditions.

The relative role of the $2\gamma$ mechanism can increase at relatively 
large momentum transfer in SL and TL regions, for the same physical reasons, 
which 
are related to the steep decreasing of the hadronic electromagnetic FFs, as 
previously discussed in [25-28] 
and more recently in \cite{Re99}.

Let us summarize our conclusions about the properties of nucleon electromagnetic 
FF 
in the time-like region.
\begin{itemize}

\item The electric FF of the proton, which can be derived from the 
$\overline{p}+p\to e^+ +e^-$ data, in the hypothesis $|G_M|=|G_E|$ or 
$G_M^{(TL)}(s)=G_M^{(SL)}(q^2)$ $(s=-q^2)$, strongly deviates from the measured 
values of $G_E^{(SL)}(q^2)$, and from asymptotic expectations.

\item The measurement of the asymmetry ${\cal A}$ of the angular dependence of 
the 
differential cross section for $\overline{p}+p\leftrightarrow e^+ +e^-$ is 
sensitive to the relative value of $G_M$ and $G_E$.

\item 
An extrapolation to high $q^2$ of the TL experimental data indicates that the 
Phr\`agmen-Lindel\"of theorem 
will be satisfied by the magnetic proton FF, only for $s(q^2)\ge 20$ 
(GeV/c)$^2$. This 
conclusion is 
nearly independent on different assumptions concerning $|G_E|$.

\item The value of the mass parameter $m_{nd}^2$, which characterizes 
the deviation of $G_M^{(TL)}$ from the dipole dependence, is comparable for the 
magnetic proton FF in TL region,  for the electric proton FF in SL region  and 
for the magnetic FF 
of the $N\to \Delta$ transition and may  indicate that this parameter 
is related to the internal hadronic electromagnetic structure.
\item The presence of a large relative phase of magnetic and electric proton FF 
in 
the TL region, if experimentally proven at relatively large momentum transfer, 
can be considered a strong  
indication that these FFs have a different 
asymptotic behavior. 
\end{itemize}

\begin{figure}
\mbox{\epsfxsize=14.cm\leavevmode \epsffile{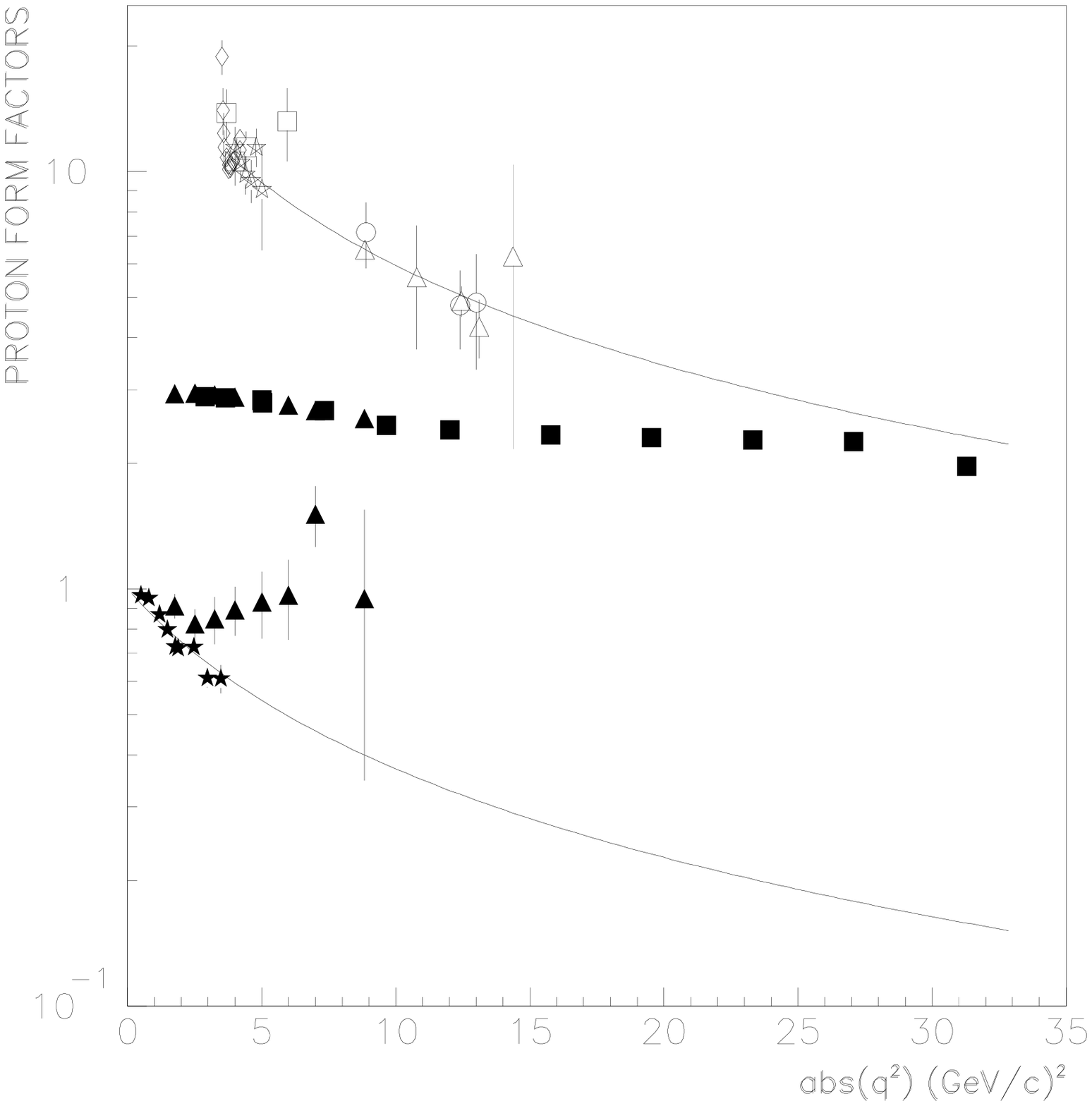}}
\caption{ Existing data for electric and magnetic form factors in space like and 
time-like regions, scaled by dipole, 
as functions of the modulus of $q^2$.
Data in space-like region are taken from: \protect\cite{Jo00} (stars), 
\protect\cite{And94} (solid triangles), \protect\cite{Ar86} (solid squares).
Data in time-like region are taken from: \protect\cite{Ar93} (open circles), 
\protect\cite{An94} (open squares), \protect\cite{Ba94} (open diamonds) 
\protect\cite{Am99} (open triangles), \protect\cite{Bi83} (open stars).
}
\end{figure}
\begin{figure}
\mbox{\epsfxsize=14.cm\leavevmode \epsffile{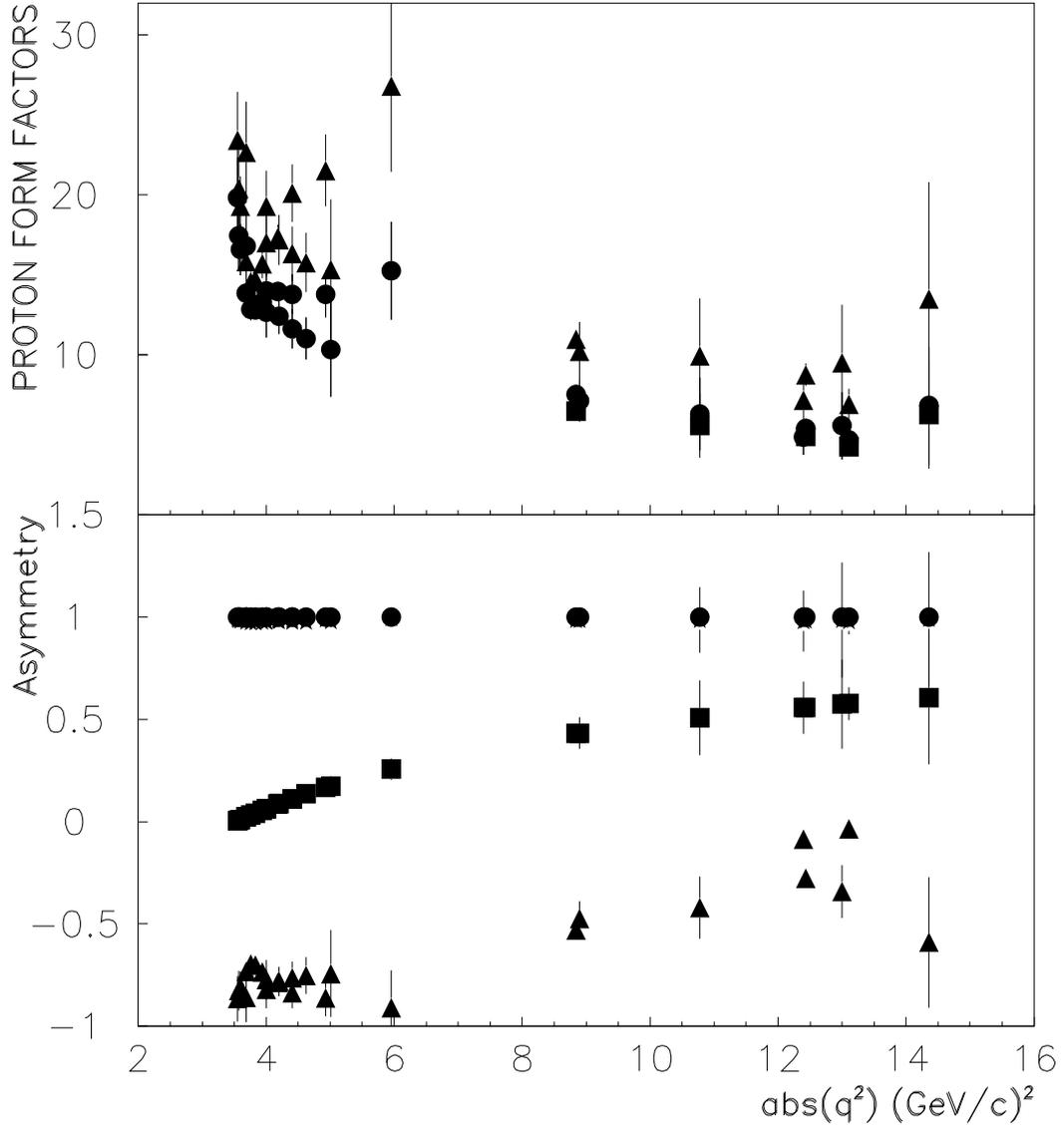}}
\caption{Nucleon form factors (top) and asymmetry (bottom) in TL region, 
deduced from the data following different assumptions (see text).
case 1: $G_E=0$ (circles); case 2: $G_E=G_M$ (squares); 
case 3: $G_M=$dipole (triangles); 
case 4: $G_E$ from \protect\cite{Jo00} (stars).}
\end{figure}
\end{document}